\documentclass[onecolumn]{article}
\usepackage[letterpaper, margin=1.5cm]{geometry}

\usepackage{cite}
\usepackage{amsmath,amssymb,amsfonts}
\usepackage{algorithmic}
\usepackage{graphicx}
\usepackage{textcomp}
\usepackage{xcolor}
\usepackage{float}
\usepackage{booktabs}
\usepackage{multirow}
\PassOptionsToPackage{hyphens}{url}\usepackage{hyperref}
\usepackage{adjustbox}
\usepackage{authblk}

\def\BibTeX{{\rm B\kern-.05em{\sc i\kern-.025em b}\kern-.08em
    T\kern-.1667em\lower.7ex\hbox{E}\kern-.125emX}}

\begin{document}

\title{Using Malware Detection Techniques \\ for HPC Application Classification}

\author[1]{Thomas Jakobsche}
\author[1]{Florina M. Ciorba}

\affil[1]{Department of Mathematics and Computer Science, University of Basel, Switzerland\\
\texttt{thomas.jakobsche@unibas.ch, florina.ciorba@unibas.ch}}

\date{}

\maketitle

%%%%%%%%%%%%%%%%%%%%%%%%%%%%%%%%%%%%%%%%%%%

\begin{abstract}
HPC systems face security and compliance challenges, particularly in preventing waste and misuse of computational resources by unauthorized or malicious software that deviates from allocation purpose. Existing methods to classify applications based on job names or resource usage are often unreliable or fail to capture applications that have different behavior due to different inputs or system noise. This research proposes an approach that uses similarity-preserving fuzzy hashes to classify HPC application executables. By comparing the similarity of SSDeep fuzzy hashes, a Random Forest Classifier can accurately label applications executing on HPC systems including unknown samples. We evaluate the Fuzzy Hash Classifier on a dataset of 92 application classes and 5333 distinct application samples. The proposed method achieved a macro f1-score of 90\% (micro f1-score: 89\%, weighted f1-score: 90\%). Our approach addresses the critical need for more effective application classification in HPC environments, minimizing resource waste, and enhancing security and compliance.
\end{abstract}

\section{Introduction} \label{sec:introduction}

\textbf{Problem Statement.}
HPC systems face significant security and compliance challenges. A major concern is the waste and misuse of computing resources through deviation from intended allocation purpose~\cite{ates2018taxonomist, peisert2010fingerprinting}. This includes the execution of software that is not within the scope of the allocation or is otherwise wasteful, disruptive, or malicious. Such software wastes computational resources, and compromises system performance and integrity.

The need for insights about executed software is highlighted by a series of cyberattacks and reports of cryptocurrency mining on HPC Systems~\cite{bbc2020europe, bbc2014us, Theodore2014harvard, Tiffany2018russian}. We adjust the guiding questions proposed in earlier work \cite{peisert2010fingerprinting} to our context of classifying the applications executed on an HPC system:
\begin{enumerate}
    \item Is an application similar or different to the applications a user or their group normally execute?
    \item Is an application similar to a (known) set of applications that are normally executed for the purpose of a particular allocation?
    \item Is an application similar to a (known) set of applications that should not be executed on the HPC system?
\end{enumerate}

Answering these questions and classifying applications that are executed on HPC systems is crucial to mitigating the above mentioned risks and ensuring the efficient and secure use of shared computational resources.

\textbf{Limitations of Existing Solutions.} 
Typical indicators of application identity include job names and application executable names~\cite{costa2021systematically}. However, these identifiers can be easily and arbitrarily changed by users~\cite{ates2018taxonomist}. User-provided job names (e.g. "my\_job") and application executable names (e.g. "a.out") are misleading if used repeatedly for different applications.

Classifying applications based on dynamic resource usage can be successful~\cite{peisert2010fingerprinting, ates2018taxonomist}. 
However, it can also introduce overhead from system monitoring and be unreliable due to system noise or previously unseen application inputs~\cite{jakobsche2021execution}. 
System noise and unseen application inputs change application behavior, resulting in varying resource usage, which in turn leads to differing classifications of the same applications~\cite{ramos2019accurate}. 
Additionally, depending on the approach, classification may only occur after the application completes execution.

Site-provided (loaded) modules can oftentimes provide clues on software usage. LMOD can track which modules are used~\cite{mclay2023lmod}. However, loaded modules cannot be reliably mapped to individual applications~\cite{peisert2010fingerprinting}.

Approaches involving static hash-based analyses of application executables concluded that cryptographic hashes can only be used to find exact matches~\cite{yamamoto2018classifying}, and proceeded to analyse file characteristics without hashing. 

\textbf{Proposed Solution.}
To reliably and accurately classify applications, in this work we use static code analysis, a method previously proposed for HPC security~\cite{haridas2020code, hpc2016security, peisert2010fingerprinting, peisert2017security}. 
Specifically, hash-based signatures of application executable files have been utilized in an HPC context for software tracking~\cite{agrawal2014user}, to collect application-specific resource usage data~\cite{yamamoto2018classifying}, and as workload identifiers in secure identity frameworks~\cite{spireXspiffe}. 
Instead of cryptographic hashes, we use similarity-preserving fuzzy hashes computed with SSDeep~\cite{kornblum2006identifying}. Fuzzy hashing for malware detection is a common approach as exemplified by Microsoft 365 Defender research~\cite{edir2021combing}.

Traditional (cryptographic) hashes can be used to match identical files and solve the problem of recognizing repeated executions of the same application~\cite{yamamoto2018classifying}. Cryptographic hashes already cover a lot of cases as users frequently execute jobs by changing the input data and not the application executable. However, cryptographic hashes fail to match application samples from the same application class when the samples differ due to small code or version changes.

Fuzzy hashing on the other hand, is a sophisticated technique that enables the comparison of similar files. This capability is particularly valuable in the context of malware detection, where variants of a single sample of malware exhibit minor modifications to evade detection mechanisms. We employ SSDeep fuzzy hashing to extract fuzzy hashes from application executable files. In our context, the SSDeep method enables us to match application samples that have slight differences in code or version changes.

We 'fuzzy hash' multiple features of application samples, such as the raw binary content, the continuous printable characters (from the \texttt{strings} command), and the global function names (extracted from the symbol table with the \texttt{nm} command). We then use a Random Forest Classifier to train a model, which we call \textit{Fuzzy Hash Classifier}, to classify application samples based on the similarity of their SSDeep fuzzy hashes.

\textbf{Expected Impact.}
The goal of this work is to avoid inefficient wasteful use and misuse of HPC systems by applications that deviate from allocation purpose and/or are otherwise disruptive, illicit, or malicious. 
Classifying application executables into known and unknown application classes will address security and compliance challenges related to executed software.
When a user suddenly executes completely different application executables within the same project allocation, this indicates a deviation from allocation purpose or even stolen access credentials. 
By tracking which specific applications are executed during resource bottlenecks (e.g. slow file system response time), the specific applications behind such issues can be pinpointed.

Labelling applications is also important for a variety of other use cases, such as reporting software usage across the cluster, analyzing performance variation of jobs~\cite{costa2021systematically}, and application-specific system optimizations such as CPU frequency tuning for known compute- vs. memory-bound applications~\cite{benkner2014automatic}.

We use a machine learning approach to classify application samples into application classes, with different fuzzy hashes as features, based on their similarity to known samples. 
Application classes represent different versions of the "same" application.
To carry out our research, we propose a \textit{Fuzzy Hash Classifier} that labels applications based on their classes or labels them as "unknown" if the application class is unknown. 

\textbf{Scope.}
The focus of this work is on assessing the feasibility of a Fuzzy Hash Classifier for application classification based on SSDeep fuzzy hashes. 
We use a dataset of preinstalled application executables (labeled based on file paths) from the production cluster
sciCORE~\cite{sciCORE} of the University of Basel, Switzerland.
This dataset does not contain malicious or illicit software.
The implementation of a dynamic collection mechanism of user executed applications is beyond the scope of this work. However, dynamic data collection could be achieved with a Slurm prolog script approach~\cite{yamamoto2018classifying}.
Automated decision making such as stopping or altering jobs is also beyond the scope of this work.

\textbf{Contribution.}
This work presents an early exploration of fuzzy hashing for application classification in an HPC context, inspired by malware detection techniques.
We evaluate the Fuzzy Hash Classifier with a dataset of 92 application classes containing 5333 distinct application samples. 
To be representative of production deployment, the train-test split includes completely unseen application classes. 
The Fuzzy Hash Classifier achieved a macro f1-score of 90\% (average f1-scores across all classes), a micro f1-score of 89\% (considering all samples equally), and a weighted f1-score of 90\% (weighing f1-scores by class size). 
This work can be extended through incorporating additional features, more sophisticated feature engineering, and using real data from a production HPC system, see Section~\ref{sec:future}.

\textbf{Organization.}
The remainder of this work is organized as follows.
We discuss related work in Section~\ref{sec:related}. 
The fuzzy hashing methodology is presented in Section~\ref{sec:methodology}. 
The results are presented in Section~\ref{sec:results} and discussed in Section~\ref{sec:discussion}.
And Section~\ref{sec:conclusion} concludes our work and outlines future work. 

\section{Related Work} \label{sec:related}
\textbf{Classification based on Resource Usage.}
A method for fingerprinting dynamic communication and computation of jobs executing on HPC systems was presented in \cite{peisert2010fingerprinting}. The fingerprints are built from IPM (Integrated Performance Monitoring) data, mainly focusing on MPI (Message Passing Interface) library calls and are used to classify the behavior of jobs.
Another approach that used classification based on resource usage was presented in \cite{ates2018taxonomist}. Different statistical features were extracted from over 700 system metrics. A machine learning approach was then used to classify application executions on individual nodes. 
A method that used performance counters to perform clustering of similar applications was presented in \cite{ramos2019accurate}. They found that the same applications are sometimes present in multiple clusters. The work concluded that this was caused by different input sizes changing application behavior and/or triggering specific parts of the code to be executed which is not executed otherwise.
A simplistic dictionary-based recognition approach was proposed in \cite{jakobsche2021execution}. This method involved building a dictionary of resource usage fingerprints of application executions inspired by the Shazam algorithm for song recognition.

The present work does not rely on monitoring data collected during execution of an application. Instead, the focus is on a static approach using features extracted from executables of applications already installed on the system.

\textbf{Classification based on Executable Analysis.}
An approach to compute the similarity of software codes was proposed in \cite{haridas2020code}. Their approach relies on a machine learning setup analyzing the CFG (Control-Flow Graph) of a code and different graph similarity measures. Extracting the CFG of an application is a complicated reverse engineering effort.
In contrast to this approach, we follow a simpler hashing-based approach.
Classification of jobs based on static information extracted from the job script and the executable file (e.g. function names) was presented in \cite{yamamoto2018classifying}. The resulting job classification is subsequently used to estimate/predict characteristics like power consumption based on jobs with similar properties.

The above work~\cite{yamamoto2018classifying} is seminal for our approach as it introduces the mechanism to collect application executables and explores cryptographic hashes of applications for classification. Cryptographic hashes are insufficient to account for variations in such as the compilation process.
This is why the above work continued with analyzing the list of raw function names through paragraph vector models. We extend this previous work~\cite{yamamoto2018classifying} by exploring the use of more flexible fuzzy hashes instead of cryptographic hashes. Using fuzzy hashes tremendously reduces the associated complexity and storage requirements of analyzing raw application executable characteristics and avoids integrity and privacy concerns of accessing raw content of users' files.

\section{Methodology} \label{sec:methodology}
\begin{figure}[H]
    \centering
    \includegraphics[width=0.4\textwidth]{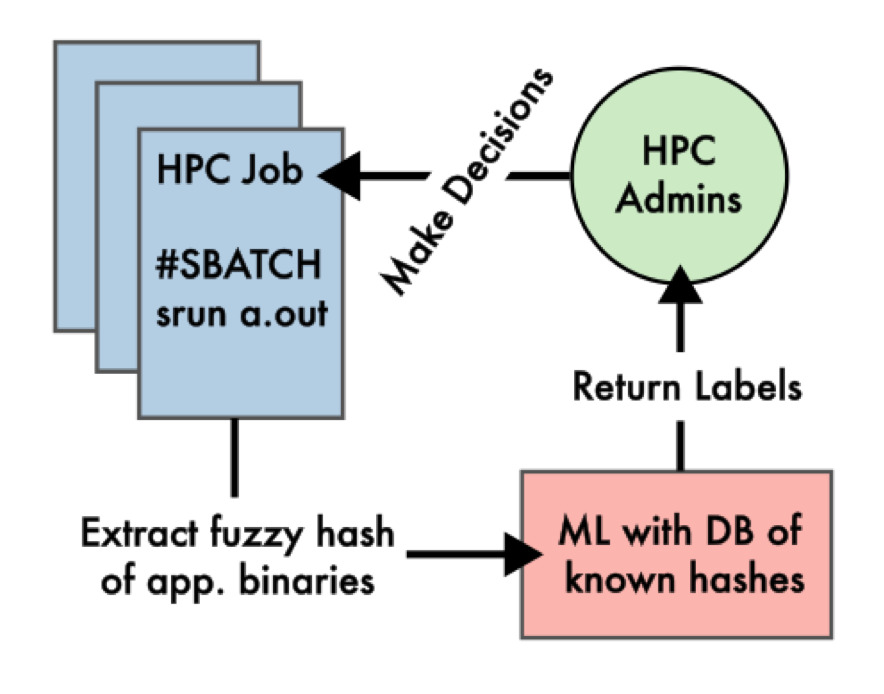}
    \caption{Proposed envisioned workflow for classifying applications and supporting decision-making about jobs.}
    \label{figure:overview}
\end{figure}

Figure~\ref{figure:overview} shows our approach and envisioned workflow. 
Fuzzy hash features are collected from applications executed inside HPC jobs.
The jobs receive an application label based on the similarity of these fuzzy hashes assessed by our 
Fuzzy Hash Classifier
approach.
Researchers and administrators can analyze and/or make decisions about HPC jobs based on these labels.

Our methodology includes data collection of preinstalled scientific applications, extracting SSDeep fuzzy hash features, and employing the Fuzzy Hash Classifier to classify application samples and return application class labels for un/known application samples. Our code is written in Python and for ML we use the Python Scikit-Learn library~\cite{pedregosa2011scikit}.

\textbf{Data Collection.}
We use python code to collect and process application sample data. 
We gather executable files from system directories that contain preinstalled software distributions 
and applications for domain science (biology, chemistry, physics, and math) from the production cluster
\mbox{sciCORE}~\cite{sciCORE} of the University of Basel, Switzerland.
The system directories with preinstalled software follow a consistent format of (e.g. OpenMalaria/46.0-iomkl-2019.01/openmalaria and OpenMalaria/43.1-foss-2021a/openmalaria), where the root folder is what we denote the application class, the sub-folders are the different versions, and within the version sub-folders are the application samples.
For each preinstalled application class, 
we traverse the version sub-folders and collect the executable files that exist in all versions and are not stripped of information (e.g.
that have an intact symbol table).

For a meaningful train-test split, we collect applications with at least 3 versions, resulting in at least 3 samples per class. Some application classes have only one sample per version sub-folder (such as the above mentioned OpenMalaria). Other application classes have multiple samples per version sub-folder, such as if the workflow of the application has multiple steps in distinct executables. Table~\ref{table:velvet} shows Velvet, a DNA sequence assembler. This application class has 2 application samples per version, velveth is used to hash DNA sequences, and velvetg is used to perform the actual assembly of the hashed sequences.

\begin{table}[H]
\caption{Versions and Executables for the Velvet Application}
\centering
\begin{tabular}{lll}
\toprule
\textbf{Class} & \textbf{Application Version} & \textbf{Samples} \\ 
\midrule
\multirow{3}{*}{Velvet} & 1.2.10-GCC-10.3.0-mt-kmer\_191 & velveth, velvetg \\ \cline{2-3} 
                        & 1.2.10-goolf-1.4.10           & velveth, velvetg \\ \cline{2-3} 
                        & 1.2.10-goolf-1.7.20           & velveth, velvetg \\ 
\bottomrule
\end{tabular}
\label{table:velvet}
\end{table}

\begin{figure}[H]
    \centering
    \includegraphics[width=0.5\textwidth]{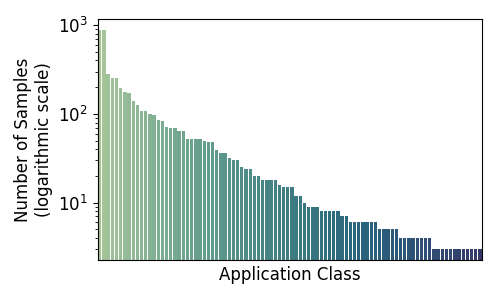}
    \caption{Number of samples for 92 application classes on a logarithmic scale.}
    \label{figure:dataset}
\end{figure}

Figure~\ref{figure:dataset} shows the application classes and their number of samples. 
There are application classes with more than one executable per version, resulting in multiple samples per version, and thus a highly imbalanced multi-class dataset. 

As our dataset is quite large, giving an overview of all application classes in one table is challenging. However, combining Table~\ref{table:classification-report} and Table~\ref{table:unknown-classes} shows all 92 application classes used for our Fuzzy Hash Classifier.

We accept this imbalance to be representative of a real world scenario, where we will also encounter significant imbalance in the frequency and diversity of application versions. 
We address class imbalance through assigning balanced weights to classes inversely proportional to class frequencies, described later with the Fuzzy Hash Classifier method. 

\textbf{Feature Extraction.}
We extract several features
of the application executable files using SSDeep fuzzy hashing, the features are fuzzy hashes of:

\begin{itemize}
    \item The raw binary content of the executable file.
    \item The continuous printable characters extracted using the \texttt{strings} command (embedded text).
    \item The global text symbols extracted using the \texttt{nm} command (function and variable names in the symbol table).
\end{itemize}

We use SSDeep to generate fuzzy hash features, which relies on Context Triggered Piecewise Hashing (CTPH). The process generates a fuzzy hash of the input in a way that allows for comparison even if the inputs are not identical. CTPH works through:
\begin{itemize}
    \item \textbf{Chunking:} SSDeep divides the input into chunks based on the content, rather than fixed-size blocks. This makes it "context triggered."
    \item \textbf{Hashing:} Each chunk is hashed using a rolling hash function, and these hashes are combined to produce a final fuzzy hash.
    \item \textbf{Comparison:} When comparing two SSDeep hashes, a similarity score is calculated based on the edit distance (specifically, the Damerau-Levenshtein (DL) distance~\cite{damerau1964technique}) between the two hashes.
    \item \textbf{Distance:} The Damerau-Levenshtein distance is a variant of the Levenshtein distance~\cite{levenshtein1966binary} 
    that counts the amount of insertions, deletions, substitutions and also transpositions of characters necessary to transform one fuzzy hash into an other.
    \item \textbf{Similarity:} SSDeep scales the DL distance into a score on a range of 0 to 100, where 0 means no similarity and 100 means that the inputs are identical.
\end{itemize}

\begin{table*}[t]
\caption{Hash Similarity Example}
\begin{adjustbox}{width=\textwidth,center}
\label{table:hash-example}
\begin{tabular}{lllrc}
\toprule
\textbf{Class} & \textbf{Version} & \textbf{Fuzzy Hash of Symbols} & \textbf{Similarity} \\
\midrule
OpenMalaria & 46.0-iomkl-2019.01 & 1536:z5ujB2ip\textcolor{blue}{prvzwz}K8l\textcolor{blue}{8lPRCuN0L830XmR8c/dGSpTWK5f5Kuy1a}zM/M3rw\textcolor{blue}{83r}w\textcolor{blue}{La6Ftl}jyx:C5ujBf\textcolor{blue}{Qz}r & \multirow{2}{*}{79} \\
OpenMalaria & 43.1-foss-2021a & 1536:3bn92z\textcolor{blue}{prvzwz}e\textcolor{blue}{8lPRCuN0L830XmR8c/dGSpTWK5f5Kuy1a}OMP\textcolor{blue}{83r}F\textcolor{blue}{La6Ftl}DJIzu:3bn9u\textcolor{blue}{Qz}Y & \\
\bottomrule
\end{tabular}
\end{adjustbox}
\end{table*}

\textbf{Damerau-Levenshtein.}
The function \( d_{a,b}(i,j) \) is defined as the Damerau–Levenshtein distance between two strings \( a \) and \( b \), whose value is the distance between an \( i \)-symbol prefix of string \( a \) and a \( j \)-symbol prefix of \( b \).

The recursive distance function is defined in Equation~\ref{eq:damerau_levenshtein}, where \( 1_{(a_i \neq b_j)} \) is the indicator function equal to 0 when \( a_i = b_j \) and otherwise equal to 1.

\begin{figure}[H]
\begin{equation}
    \label{eq:damerau_levenshtein}
    d_{a,b}(i,j) = 
    \min \begin{cases}
    0 & \text{if } i=j=0, \\
    d_{a,b}(i-1,j) + 1 & \text{if } i > 0, \\
    d_{a,b}(i,j-1) + 1 & \text{if } j > 0, \\
    d_{a,b}(i-1,j-1) + 1_{(a_i \neq b_j)} & \text{if } i,j > 0, \\
    d_{a,b}(i-2,j-2) + 1_{(a_i \neq b_j)} & \text{if } i,j > 1 \text{ and } a_i = b_{j-1} \text{ and } a_{i-1} = b_j
    \end{cases}
\end{equation}
\end{figure}

Below are the cases covered by the Damerau–Levenshtein distance:

\begin{itemize}
    \item \( d_{a,b}(i-1,j) + 1 \), a deletion (from \( a \) to \( b \)),
    \item \( d_{a,b}(i,j-1) + 1 \), an insertion (from \( a \) to \( b \)),
    \item \( d_{a,b}(i-1,j-1) + 1_{(a_i \neq b_j)} \), a match or mismatch, depending on whether the symbols are the same,
    \item \( d_{a,b}(i-2,j-2) + 1_{(a_i \neq b_j)} \), a transposition between two successive symbols.
\end{itemize}

The Damerau–Levenshtein distance between \( a \) and \( b \) is then given by the function value for the full strings: \( d_{a,b}(|a|,|b|) \), where \( i = |a| \) is the length of string \( a \), and \( j = |b| \) is the length of string \( b \).

The use of CTPH and the Damerau-Levenshtein
distance allows SSDeep to detect similarities between inputs even when they are not identical, which is useful for malware detection and HPC executable classification (our context). 
We compute a feature matrix for our dataset based on the SSDeep fuzzy hash similarity between sample features. 
Table~\ref{table:hash-example} shows an example of the comparison of two hashes from two different versions of the OpenMalaria application class. Identical common sub-strings are highlighted in blue.
The specific Damerau-Levenshtein distance is more complex than just the common sub-strings.
However, this exemplifies the principle between comparing fuzzy hashes. 

The resulting SSDeep similarity between two fuzzy hashes is a useful measure for quantifying the degree of similarity or variation between files. This is critical for identifying near-duplicates or minor modifications in digital forensics and malware analysis. The SSDeep similarity captures structural similarities even when files differ slightly. This also reduces the overhead of comparing large-scale datasets because it avoids exhaustive and costly byte-by-byte comparison of files.

\textbf{Fuzzy Hash Classifier.}
We implement a two-phase train-test split using Scikit-Learn library~\cite{pedregosa2011scikit}. 
In the first phase we split the application classes in a 80-20 train-test manner into known and unknown classes to ensure we have completely unknown application samples in our test set. 
In the second phase we further split the known classes
through a stratified 60-40 train-test split on the samples. 
The two-phase train-test approach splits the initial 5333 samples into a training set of 2688 samples and a test set of 2645 samples (including 852 samples of completely unknown classes).
We employ supervised learning, specifically classification, using the Random Forest Classifier algorithm to build our Fuzzy Hash Classifier. 
We optimize the performance of the Fuzzy Hash Classifier with hyperparameter tuning through grid search only within the training set.

We address the class imbalance through assigning balanced weights to the classes where the weights are inversely proportional to class frequencies. This is a common practice to ensure that a ML model does not favor the majority classes. 
We chose the Random Forest Classifier for two reasons: 

\begin{itemize}
    \item \textbf{Non-linearity:} Random Forests capture complex, non-linear relationships between features and the target variable. Random Forests are thus suitable for our feature matrix that consists of abstract fuzzy hash similarity values and not explicit distances in Euclidean space.
    \item \textbf{Feature Importance:} Random Forests provide feature importance scores, giving insights into which features are most influential in the predictions.
\end{itemize}

\textbf{Unknown Samples.}
Table~\ref{table:unknown-classes} shows the application classes that were randomly put into the class of unknown samples through the 80-20 train-test split. 
The class of unknown samples is important to test model robustness, because in a production setting the model will encounter samples from application classes that it has never seen before, and needs to accurately label them as "unknown".
"Unknown" in this context means that a sample might be deviating from allocation purpose because it does not fit into known categories. A known category could be the "usual" applications executed by a user, and then encountering an "unknown" application indicates that the user is suddenly executing something else.
In the Fuzzy Hash Classifier the label of the "unknown" class is "-1".

\begin{table}[H]
\caption{Class of Unknown Samples}
\label{table:unknown-classes}
\begin{adjustbox}{width=0.3\textwidth,center}
\begin{tabular}{lr}
\toprule
\textbf{Application Class} & \textbf{Sample Count} \\
\midrule
Schrodinger        & 195 \\
QuantumESPRESSO    & 178 \\
SAMtools           & 108 \\
MCL                & 52  \\
BLAST              & 52  \\
FASTA              & 48  \\
MolProbity         & 39  \\
AUGUSTUS           & 36  \\
HISAT2             & 30  \\
OpenMalaria        & 25  \\
Gurobi             & 20  \\
Kraken             & 18  \\
METIS              & 18  \\
CCP4               & 9   \\
TM-align           & 9   \\
ClustalW2          & 4   \\
dssp               & 4   \\
libxc              & 4   \\
CHARMM             & 3   \\
\bottomrule
\end{tabular}
\end{adjustbox}
\end{table}

\textbf{Classification.}
The Fuzzy Hash Classifier returns class labels for application executables (samples) based on their similarity to other known samples in our dataset.

\begin{itemize}
    \item \textbf{Classifying Applications:} The Fuzzy Hash Classifier labels test samples based on the similarity of their fuzzy hash features compared to training samples.
    \item \textbf{Labeling Unknown Samples:} Samples not similar to any other known samples are labeled as "unknown", indicating it is a new or unseen sample. This classification is based on a confidence threshold applied to the probability prediction of the Fuzzy Hash Classifier. The confidence threshold is tuned as part of the hyperparameter grid search within the training set. 
\end{itemize}

\textbf{Evaluation.}
We use the micro, macro, and weighted versions of precision, recall, and f1-score~\cite{van1979information} for evaluation.
\textit{Precision} quantifies how many of the positive predictions were actually correct. 
It is calculated as the ratio of true positives to the sum of true positives and false positives.
\textit{Recall} (or sensitivity) quantifies how many of the actual positives were correctly identified.
Recall is calculated as the ratio of true positives to the sum of true positives and false negatives.
The \textit{f1-score} is the harmonic mean of precision and recall, shown in Equation~\ref{eq:f1-score}. It balances the two metrics, especially when there is an uneven class distribution, giving a single measure.

\begin{equation}
    \label{eq:f1-score}
    f_1 = 2 \cdot \frac{\text{Precision} \cdot \text{Recall}}{\text{Precision} + \text{Recall}}
\end{equation}

\textit{Micro-averaging} aggregates the contributions of all classes, treating all instances equally. 
\textit{Macro-averaging} calculates metrics for each class independently and then takes their unweighted average, giving equal importance to each class. 
\textit{Weighted-averaging}, on the other hand, adjusts for class imbalances by averaging metrics with weights proportional to the number of true instances in each class, providing a more representative performance measure in imbalanced datasets.

\section{Results} \label{sec:results}
\textbf{Classification Report.}
Table~\ref{table:classification-report} shows the classification report generated by the Scikit-Learn library.
The micro average for precision, recall, and f1-score are identical, in this case they correspond to the overall accuracy. 
This is because the micro averages consider the total true positives, false positives, and false negatives across all classes, effectively calculating the metrics as if it were a binary classification, which is equivalent to the overall accuracy in a multi-class context.

\begin{table}[H]
\caption{Classification Report}
\label{table:classification-report}
\begin{adjustbox}{width=0.4\textwidth,center}
\begin{tabular}{lcccc}
\toprule
\textbf{Class} & \textbf{Precision} & \textbf{Recall} & \textbf{f1-Score} & \textbf{Support} \\
\midrule
-1                & 0.92 & 0.75 & 0.83 & 852 \\
Augustus          & 0.42 & 1.00 & 0.59 & 10 \\
BCFtools          & 1.00 & 1.00 & 1.00 & 4 \\
BEDTools          & 1.00 & 1.00 & 1.00 & 3 \\
BLAT              & 1.00 & 1.00 & 1.00 & 5 \\
BWA               & 1.00 & 0.40 & 0.57 & 5 \\
BamTools          & 1.00 & 0.50 & 0.67 & 2 \\
BigDFT            & 0.55 & 0.96 & 0.70 & 28 \\
CAD-score         & 0.25 & 1.00 & 0.40 & 3 \\
CD-HIT            & 1.00 & 1.00 & 1.00 & 12 \\
CapnProto         & 1.00 & 1.00 & 1.00 & 1 \\
Cas-OFFinder      & 1.00 & 1.00 & 1.00 & 1 \\
Celera Assembler  & 1.00 & 0.96 & 0.98 & 101 \\
Cell-Ranger       & 0.39 & 0.89 & 0.54 & 28 \\
CellRanger        & 0.79 & 0.95 & 0.86 & 20 \\
Cufflinks         & 0.60 & 0.50 & 0.55 & 6 \\
DIAMOND           & 1.00 & 0.50 & 0.67 & 2 \\
Exonerate         & 1.00 & 0.91 & 0.95 & 43 \\
FSL               & 1.00 & 1.00 & 1.00 & 351 \\
FastTree          & 1.00 & 1.00 & 1.00 & 2 \\
GMAP-GSNAP        & 1.00 & 1.00 & 1.00 & 38 \\
HH-suite          & 0.74 & 0.96 & 0.83 & 26 \\
HMMER             & 0.97 & 1.00 & 0.99 & 34 \\
HTSlib            & 0.40 & 0.33 & 0.36 & 6 \\
Infernal          & 1.00 & 1.00 & 1.00 & 7 \\
InterProScan      & 0.75 & 0.99 & 0.86 & 102 \\
JAGS              & 1.00 & 1.00 & 1.00 & 1 \\
Jellyfish         & 1.00 & 1.00 & 1.00 & 2 \\
Kraken2           & 1.00 & 1.00 & 1.00 & 6 \\
MAGMA             & 1.00 & 1.00 & 1.00 & 1 \\
MATLAB            & 0.70 & 1.00 & 0.82 & 14 \\
MMseqs2           & 1.00 & 1.00 & 1.00 & 1 \\
MUMmer            & 0.95 & 0.69 & 0.80 & 26 \\
Mash              & 1.00 & 1.00 & 1.00 & 1 \\
MolScript         & 1.00 & 1.00 & 1.00 & 3 \\
MrBayes           & 1.00 & 1.00 & 1.00 & 1 \\
OpenBabel         & 1.00 & 1.00 & 1.00 & 8 \\
OpenMM            & 1.00 & 1.00 & 1.00 & 2 \\
OpenStructure     & 1.00 & 1.00 & 1.00 & 56 \\
PLUMED            & 1.00 & 1.00 & 1.00 & 3 \\
PRANK             & 1.00 & 1.00 & 1.00 & 2 \\
PSIPRED           & 1.00 & 1.00 & 1.00 & 7 \\
PhyML             & 1.00 & 1.00 & 1.00 & 2 \\
RECON             & 1.00 & 1.00 & 1.00 & 6 \\
RSEM              & 0.95 & 0.86 & 0.90 & 21 \\
Racon             & 1.00 & 1.00 & 1.00 & 2 \\
Raster3D          & 1.00 & 1.00 & 1.00 & 13 \\
RepeatScout       & 1.00 & 1.00 & 1.00 & 2 \\
Rosetta           & 0.66 & 0.85 & 0.75 & 114 \\
SMRT-Link         & 1.00 & 0.67 & 0.80 & 3 \\
SOAPdenovo2       & 1.00 & 1.00 & 1.00 & 2 \\
STAR              & 1.00 & 1.00 & 1.00 & 10 \\
Salmon            & 1.00 & 0.33 & 0.50 & 3 \\
SeqPrep           & 1.00 & 1.00 & 1.00 & 3 \\
Stacks            & 0.99 & 0.99 & 0.99 & 69 \\
StringTie         & 1.00 & 0.50 & 0.67 & 2 \\
Subread           & 1.00 & 1.00 & 1.00 & 21 \\
TopHat            & 0.66 & 1.00 & 0.79 & 19 \\
Trinity           & 1.00 & 1.00 & 1.00 & 41 \\
VCFtools          & 1.00 & 1.00 & 1.00 & 2 \\
VSEARCH           & 1.00 & 1.00 & 1.00 & 1 \\
Velvet            & 1.00 & 1.00 & 1.00 & 2 \\
ViennaRNA         & 0.96 & 0.93 & 0.95 & 29 \\
XDS               & 0.63 & 1.00 & 0.77 & 34 \\
breseq            & 1.00 & 1.00 & 1.00 & 4 \\
canu              & 1.00 & 1.00 & 1.00 & 51 \\
cdbfasta          & 1.00 & 1.00 & 1.00 & 2 \\
fastQValidator    & 1.00 & 1.00 & 1.00 & 2 \\
fastp             & 1.00 & 1.00 & 1.00 & 1 \\
fineRADstructure  & 1.00 & 1.00 & 1.00 & 2 \\
kallisto          & 1.00 & 0.50 & 0.67 & 2 \\
kentUtils         & 1.00 & 0.99 & 0.99 & 352 \\
prodigal          & 1.00 & 1.00 & 1.00 & 1 \\
segemehl          & 1.00 & 1.00 & 1.00 & 1 \\
\bottomrule
\textbf{micro avg}       & \textbf{0.89} & \textbf{0.89} & \textbf{0.89} & \textbf{2645} \\
\textbf{macro avg}      & \textbf{0.92} & \textbf{0.92} & \textbf{0.90} & \textbf{2645} \\
\textbf{weighted avg}   & \textbf{0.92} & \textbf{0.89} & \textbf{0.90} & \textbf{2645} \\
\hline
\end{tabular}
\end{adjustbox}
\end{table}

The \textit{Support} column in the classification report (Table~\ref{table:classification-report}) denotes the number of samples used for testing. Not all classes are represented in the classification report because some classes are grouped under the "-1" unknown class label (see Table~\ref{table:unknown-classes}).

\textbf{Feature Importance.}
Table~\ref{table:feature-importance} shows the normalized feature importance extracted from the Random Forest Classifier. Importance scores are values assigned to each feature by the Random Forest Classifier that indicate how influential each feature is in making predictions. 
The raw importance scores are in no specific format and represent internal scores of the machine learning model that are not normalized nor scaled. For readability purposes we normalize the scores to 1.

\begin{table}[H]
    \centering
    \caption{Feature Importance (normalized)}
    \begin{adjustbox}{width=0.3\textwidth,center}
    \begin{tabular}{@{}lcr@{}}
    \toprule
    \textbf{Features} & \textbf{Importance} \\ \midrule
    ssdeep-file      & 0.0718            \\
    ssdeep-strings   & 0.1404            \\
    ssdeep-symbols   & 0.7879            \\ \bottomrule
    \end{tabular}
    \label{table:feature-importance}
    \end{adjustbox}    
\end{table}

\begin{figure}[H]
    \centering
    \includegraphics[width=0.6\textwidth]{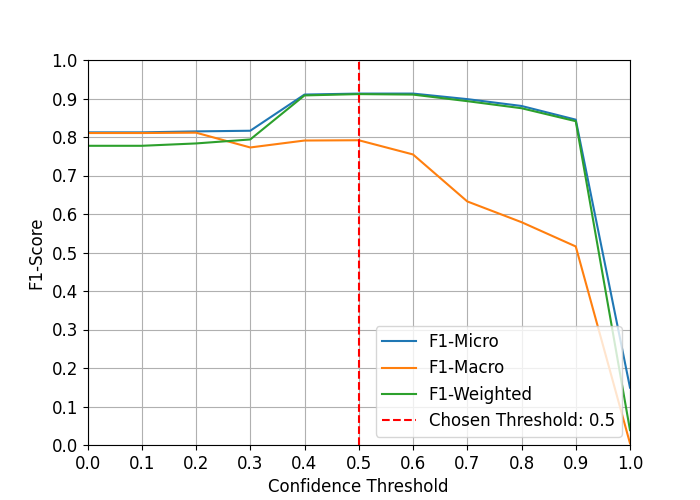}
    \caption{The f1-Score over confidence threshold of the grid search within the training set to handle unknown classes.}
    \label{figure:ml-tuning}
\end{figure}

\textbf{Confidence Threshold.}
The confidence threshold is applied to the probability prediction of the Fuzzy Hash Classifier and decides whether the classifier predicts a specific application class or the "-1" unknown class.
Figure~\ref{figure:ml-tuning} shows the individual micro, macro, and weighted f1-scores per confidence threshold reported during the hyperparameter grid search evaluated only within the training set. 
We choose the confidence threshold that maximizes the combined micro, macro, and weighted f1-scores. The hyperparameter tuning also involved other standard parameters of the Random Forest Classifier 
(such as, n\_estimators, criterion, max\_depth, min\_samples\_split, min\_samples\_leaf, and max\_features).

\section{Discussion} \label{sec:discussion}

\textbf{Unknown Class.}
Some applications are deliberately in the "-1" unknown class to test the robustness of the model (see Table~\ref{table:unknown-classes}).
A precision value higher than recall shows that our model confidently labels a sample as "unknown" and is usually correct.
However, it fails to capture all cases of "unknown" samples, leaving some of them undetected. 
If the priority is to capture as many unknown samples as possible, 
which potentially deviate from allocation purpose because they are not among the known application classes,
then we can manually increase the confidence threshold to capture all unknown samples. However, this comes at the cost of lower precision when labeling a sample as unknown.

\textbf{Imbalanced Dataset.}
Some classes like FSL, Celera Assembler, and kentUtils have high support values (i.e. 351, 101, and 352, respectively), while others have very low support values (i.e. CapnProto, JAGS, with only 1 instance), as shown in Table~\ref{table:classification-report}. The support value reflects the number of samples of this application class in our test set. This imbalance can skew the micro and weighted metrics for underrepresented classes. 
To address dataset imbalance, we employ balanced class weighting as described in Section~\ref{sec:methodology} and report the \textit{macro f1-score} which is the average of the performance of all classes and not influenced by class imbalances. 
Application classes with a low support value represent applications that have only very few versions or even only exist as a single instance without other versions. 
In production, such applications can be recognized by a direct match of their hash if they are executed again (because we ever only encounter one version of the executable). 
In our machine learning context, application classes with few samples provide limited statistical significance.
However, we refrain from artificially including or excluding any classes, and leave the dataset imbalanced on purpose to explore and understand how our Fuzzy Hash Classifier performs.

\textbf{Inconsistent Performance.}
Certain classes like BigDFT (precision: 0.55, recall: 0.96, support: 28) and MUMmer (precision: 0.95, recall: 0.69, support: 26) exhibit significant discrepancies between precision and recall, see Table~\ref{table:classification-report}. This indicates that our model does not generalize as well across all classes. This might be due to application-specific characteristics, where certain applications change more drastically across versions than others. Analyzing these misclassifications can improve the model by highlighting a need for more diverse data, more features to distinguish samples, or both.

\textbf{CellRanger vs. Cell-Ranger.}
This is a case of two classes with a very similar name, which are indeed the same application class, see Table~\ref{table:classification-report}. However, in the directory wherefrom we scraped the application executables, different versions of CellRanger/Cell-Ranger where installed at two different locations. 
"Cell-Ranger" contains versions 2.1.1, 3.0.0, and 3.1.0, while "CellRanger" contains versions 4.0.0, 5.0.0, 6.0.1, 6.1.2, and 7.1.0. These classes are misclassified against each other, skewing the results for both classes. 

\textbf{Augustus vs. AUGUSTUS.}
As shown in Table~\ref{table:classification-report} and Table~\ref{table:unknown-classes} we also have two instances of the same class split across the labeled and unknown portion of our data. This is again due to different versions being installed in different locations. Many cases of the manually labeled unknown "AUGUSTUS" are misclassified as the known "Augustus", which is an understandable misclassification.
However, this misclassification skews the results for both the Augustus class and the "-1" unknown class. 

\textbf{Overall Performance.}
Despite significant class imbalances and certain misclassification due to inconsistently labeled data, Table~\ref{table:classification-report} shows that the micro, macro, and weighted f1-scores around 90\% are solid, indicating high overall model performance. 
However, given the high variability in class performance (specifically small classes), these results could be misleading if interpreted without considering the imbalance and individual class performance. 

\textbf{Confidence Threshold.}
Figure~\ref{figure:ml-tuning} shows the selection and performance of the confidence threshold. We have a large, manually labeled "-1" unknown class to check for robustness. As we increase the confidence threshold, the macro f1-score decreases, while the micro and weighted f1-scores remain high. This is because the high number of unknown samples significantly impacts the scores, and these unknown samples perform better with higher confidence thresholds. However, the performance of all other classes suffers with increasing confidence threshold, as shown by (and why we report) the macro f1-score.

\textbf{Feature Importance.}
Figure~\ref{table:feature-importance} shows the feature importance extracted from the Fuzzy Hash Classifier. 
Function names from the symbol table are more important than printable strings or the raw file content.
Application modification can introduce changes in both the code and the compiler, affecting features differently:
\begin{itemize}
    \item Raw file content: Changes with every modification, either due to code updates or different compiler versions or flags, as the compiler translates the code into machine-readable form.
    \item Printable strings: Frequently affected by code changes (e.g., bug fixes), as strings are directly tied to the code.
    \item Function names: Typically stable across versions unless there’s a major refactor or new functionality, as the overall structure of the code remains consistent.
\end{itemize}

\textbf{Labeled Data.}
Our dataset consists of pre-installed application samples. Our approach is equally applicable to user-compiled executables. The difference is that user-provided names are not trustworthy, because users can choose undescriptive names, such as \texttt{a.out}.
When dealing with user-compiled executables our approach can be used for 2 strategies:
\begin{itemize}
    \item Compare user-compiled executables with pre-installed applications, because users are potentially executing a different version of existing software.
    \item Compare user-compiled executables among themselves, to find if users execute what they usually execute.
\end{itemize}

By comparing user-compiled executables to pre-installed applications and the applications that are usually executed by users, our approach detects deviations and outliers that differ from commonly used software.
This classification aligns with our initial guiding questions (see Section~\ref{sec:introduction}) aiming to distinguish "regular" vs. "unusual" application classes.

\textbf{Limitations.}
Our current approach cannot distinguish different codes that are executed through wrapper scripts, such as bash or Python.
These types of software dynamically load modules and execute code at execution time.
Understanding and classifying this type of software is more complex, requiring more invasive analysis techniques, and is part of immediate future work.
Our approach also does not work with executables that have been stripped of the symbol table. Binary stripping is a method commonly used to obscure binary characteristics or save storage space. Without this information (e.g. the symbol table), our approach is unable to extract meaningful fuzzy hashes.
Stripped binaries could be dealt with by investigating other characteristics, such as the dynamic call tree of the functions. A dynamic call tree would give insight into a program's internal function call logic, providing similar function-based information as the symbol table.

\textbf{Overall assessment.}
Our model provides solid results, yet with room for improvement.
The approach can be fine-tuned for specific purposes, such as increasing the confidence threshold to capture more potentially unknown samples. 
The current results indicate that fuzzy hash-based application classification effectively label applications, providing valuable insights into what users are executing. 
Our approach answers the initial guiding questions, by reasonably assessing whether users preserve the versions of their usual application software or suddenly execute completely different application(s). 

\section{Conclusion and Future Work} \label{sec:conclusion} \label{sec:future}
\textbf{Conclusion.}
This work presents an exploration of fuzzy hashing to classify applications in HPC. 
This approach was inspired by malware detection techniques and encompasses a machine learning approach using the Random Forest Classifier and SSDeep fuzzy hashes as features. 
Our dataset includes 92 application classes with 5`333 distinct application samples. 
The Fuzzy Hash Classifier achieved a macro f1-score of 90\% (micro f1-score: 89\%, weighted f1-score: 90\%). 
This work shows that static analysis of fuzzy hash features is a promising approach, which can precede and/or complement dynamic classification approaches based on application resource usage.
This work provides insights into the topic of application classification for HPC systems and opens new avenues for future work. 

\textbf{Future Work.}
There is room for improvement in extending the number of fuzzy hash features used.
Future work could study loading shared objects extracted through the \texttt{ldd} command~\cite{yamamoto2018classifying}. Other machine learning models can also be explored and compared, such as Support Vector Machines and K-Nearest Neighbors. 
Another promising avenue for future work is to combine static binary analysis with analysis of dynamic execution behavior, which can complement each other.
Lastly, we plan to deploy and test our solution in a production system, in order to test, verify, and validate it on user-provided software.

\section*{Acknowledgment}
This work was supported by sciCORE (\url{http://scicore.unibas.ch/}), the scientific computing center at the University of Basel, Switzerland, through access to data, computational resources, and their dedicated support for this research.

\bibliographystyle{plain}
\bibliography{references.bib}

\end{document}